\documentclass[aps,prc,reprint,groupedaddress]{revtex4-1}

\usepackage{amsmath}
\usepackage{graphicx}

\newcommand{\iso}[2]{$^{#1}$#2}
\newcommand{\degree}{$^{\circ}$ }

\newcommand{\ratiocl}{$^{36}$Cl/Cl }
\newcommand{\sci}[2]{$#1 \times 10^{-#2}$ }

\newcommand{\cl}{$^{36}$Cl }

\begin{document}


\title{Measurement of the \3S($\alpha$,p)\6Cl cross section: Implications for production of \6Cl in the early Solar System}

\author{M. Bowers}
\email[]{mbowers2@nd.edu}
\affiliation{University of Notre Dame, Notre Dame, IN 46656, USA}
\author{Y. Kashiv}
\affiliation{University of Notre Dame, Notre Dame, IN 46656, USA}
\author{W. Bauder}
\affiliation{University of Notre Dame, Notre Dame, IN 46656, USA}
\author{M. Beard}
\affiliation{University of Notre Dame, Notre Dame, IN 46656, USA}
\author{P. Collon}
\affiliation{University of Notre Dame, Notre Dame, IN 46656, USA}
\author{W. Lu}
\affiliation{University of Notre Dame, Notre Dame, IN 46656, USA}
\author{K. Ostdiek}
\affiliation{University of Notre Dame, Notre Dame, IN 46656, USA}
\author{D. Robertson}
\affiliation{University of Notre Dame, Notre Dame, IN 46656, USA}

\date{\today}

\begin{abstract}
Short-lived radionuclides (SLRs) with lifetimes $\tau$ $<$ 100 Ma are known to have been extant when the Solar System formed over 4.5 billion years ago. Identifying the sources of SLRs is important for understanding the timescales of Solar System formation and processes that occurred early in its history. Extinct \6Cl (t$_{1/2}$ = 0.301 Ma) is thought to have been produced by interaction of solar energetic particles (SEPs), emitted by the young Sun, with gas and dust in the nascent Solar System. However, models that calculate SLR production in the early Solar System (ESS) lack experimental data for the \6Cl production reactions. We present here the first measurement of the cross section of one of the main \6Cl production reactions, \3S($\alpha$,p)\6Cl, in the energy range 0.70 - 2.42 MeV/A. The cross section measurement was performed by bombarding a target and collecting the recoiled \cl atoms produced in the reaction, chemically processing the samples, and measuring the $^{36}$Cl/Cl ratio of the activated samples with accelerator mass spectrometry (AMS). The experimental results were found to be systematically higher than the cross sections used in previous local irradiation models and other Hauser-Feshbach calculated predictions. However, the effects of the experimentally measured cross sections on the modeled production of \6Cl in the early Solar System were found to be minimal. Reactions channels involving S targets dominate \6Cl production, but the astrophysical event parameters can dramatically change each reactions' relative contribution.
\end{abstract}


\maketitle

\section{Introduction}
\label{sec:intro}

The presence of \iso{36}{Cl} (t$_{1/2}$ = 0.301 Ma) in the early Solar System (ESS) is detected in chondrules and Ca- Al- rich inclusions (CAIs) found in carbonaceous chondrites~\cite{Lin2005,Hsu2006,Ushikubo2007,Jacobsen2011}. Although now extinct, evidence that \6Cl was extant in the ESS is inferred from correlations between excess of its daughter, \iso{36}{S}, and Cl/S ratios in secondary alteration Cl-rich minerals (e.g., sodalite, wadalite), with the highest initial ratio measured at (\iso{36}{Cl}/\iso{35}{Cl})$_0$ = (1.81$\pm$0.13) $\times$ 10$^{-5}$~\cite{Jacobsen2011}. CAIs and chondrules were among the first solids to condense in the Solar System. Absolute Pb-Pb dating techniques from the decay of long-lived radioisotopes (e.g., \iso{238}{U} and \iso{235}{U}) have determined the age of these inclusions to be greater than 4.564 Ga~\cite{Connelly2012}. Along with \6Cl, there is experimental meteoritic evidence of other short-lived radionuclides (SLRs) that were present during the early formation of the Solar System~\cite{Wasserburg2006}. These SLRs, including \iso{10}{Be}, \iso{26}{Al}, \iso{41}{Ca}, \iso{60}{Fe}, have measured abundances over what is predicted from galactic steady-state enrichment and require nucleosynthesis shortly before or after the collapse of the Sun's parent molecular cloud~\cite{Huss2009}. Since SLR half-lives, $\leq$ 80 Ma, are very short relative to the age of the Solar System (SS), they can be used as chronometers of SS formation and early evolution. 

Inferring the source of SLRs is a complicated issue. Production of SLRs was proposed by stellar nucleosynthesis in supernovae (SNe)~\cite{Takigawa2008}, AGB stars~\cite{Wasserburg1994,Wasserburg2006} or Wolf-Rayet stars~\cite{Arnould2006}. In this scenario, the freshly-synthesized radioactivities were injected within a million years of when the Solar System began to form. However, energetic particle irradiation in the Solar Nebula or in the ESS has also been suggested as a viable origin source of SLRs~\cite{Shu1997}. With current understanding, no one source is capable of producing all SLRs at the inferred ESS abundances. This implies that SLRs were produced by multiple sources. Although \iso{36}{Cl} can be produced in AGB stars and SNe, nucleosynthetic models are unable to reproduce the measured (\iso{36}{Cl}/\iso{35}{Cl})$_0$ ratio with the other SLRs simultaneously~\cite{Wasserburg2006,Lugaro2012}. More likely, \iso{36}{Cl} was produced in the ESS during the Sun's T Tauri phase by solar energetic particle (SEPs) irradiation (mainly p, $\alpha$, \iso{3}{He})~\cite{Hsu2006,Jacobsen2011}.

The x-wind model is a framework that was proposed to explain the production of SLRs through local irradiation~\cite{Shu1997}. In this model, SEPs are accelerated from the protoSun during intense x-ray events to energies in excess of 1 MeV/A. Numerous attempts have been made to reproduce the initial Solar System \iso{36}{Cl} abundance using the x-wind model~\cite{Goswami2001,Leya2003,Gounelle2006,Sahijpal2007}. However, due to a lack of experimental cross section data the models are forced to rely on theoretical predictions, which is a large source of uncertainty in the results~\cite{Leya2003,Gounelle2006}. It was our motivation to measure the cross section for an important reaction in \6Cl production as well as identify any other important reactions that have not been previously measured.

The energy spectrum of SEPs can be modeled by a power-law distribution $\propto E^{-p}$, where $E$ is the proton energy in MeV/A and $p$ is usually between 2 and 5. Therefore the reactions with excitation functions that peak at lower energies should dominate production of the \6Cl.  The \3S($\alpha$,p)\6Cl reaction was previously shown to have a large cross section that peaks at lower energies than other production reactions considered in the irradiation models~\cite{Goswami2001,Gounelle2006}. For this reason we chose to measure the \3S($\alpha$,p)\6Cl reaction cross section. Initial results were recently published~\cite{Bowers2012}.

We present the experimental results of the \3S($\alpha$,p)\6Cl reaction cross section. The experimental results are then compared to the theoretical cross sections used in the previous irradiation models as well as Hauser-Feshbach predictions. The effects of the measured cross sections on \6Cl production were tested using previously accepted astrophysical parameters for the solar flare events. Finally, to test our relatively simple assumption that  \3S($\alpha$,p)\6Cl is the dominant production channel for \6Cl, we investigated contributions from competing \6Cl production channels and how those contributions change depending on the environment of the flare event considered.

\section{Experimental Procedure}
\label{sec:exp_procedure}

The \iso{33}{S}($\alpha$,p)\cl reaction cross section was measured at 6 energies that ranged from 0.70 to 2.42 MeV/A. The range is within the Solar flare energy spectrum~\cite{Reames1997,Feigelson2002}. The measurement was performed in three stages. Initially a \iso{4}{He} gas cell target was bombarded by a \iso{33}{S} beam. The forward-recoiled \cl atoms were collected in a catcher foil during the activations. The implanted \cl atoms were chemically extracted and mixed with a natural chlorine carrier for cathode preparation. Lastly, the \ratiocl ratios of the activated samples were measured by accelerator mass spectrometry (AMS). The technique had been previously used in studying the \iso{40}{Ca}($\alpha$,$\gamma$)\iso{44}{Ti} reaction \cite{Nassar2006,RobertsonThesis}. The inverse kinematic approach of producing \cl via the $\alpha$(\iso{33}{S},\iso{36}{Cl})p reaction had the advantage of using an isotopically-pure \iso{33}{S} beam with a high purity \iso{4}{He} target. The 0.75\% abundance of \iso{33}{S} limited beam output but avoided the complexities of using a \3S-enriched solid sulfur target, \6Cl production by competing reactions, and beam current limitations due to target degradation. Another advantage of this technique was the target thickness could be easily controlled by monitoring the pressure inside the gas cell.

\subsection{Activations}
\label{sec:activations}

\begin{figure}
\includegraphics[width = \columnwidth]{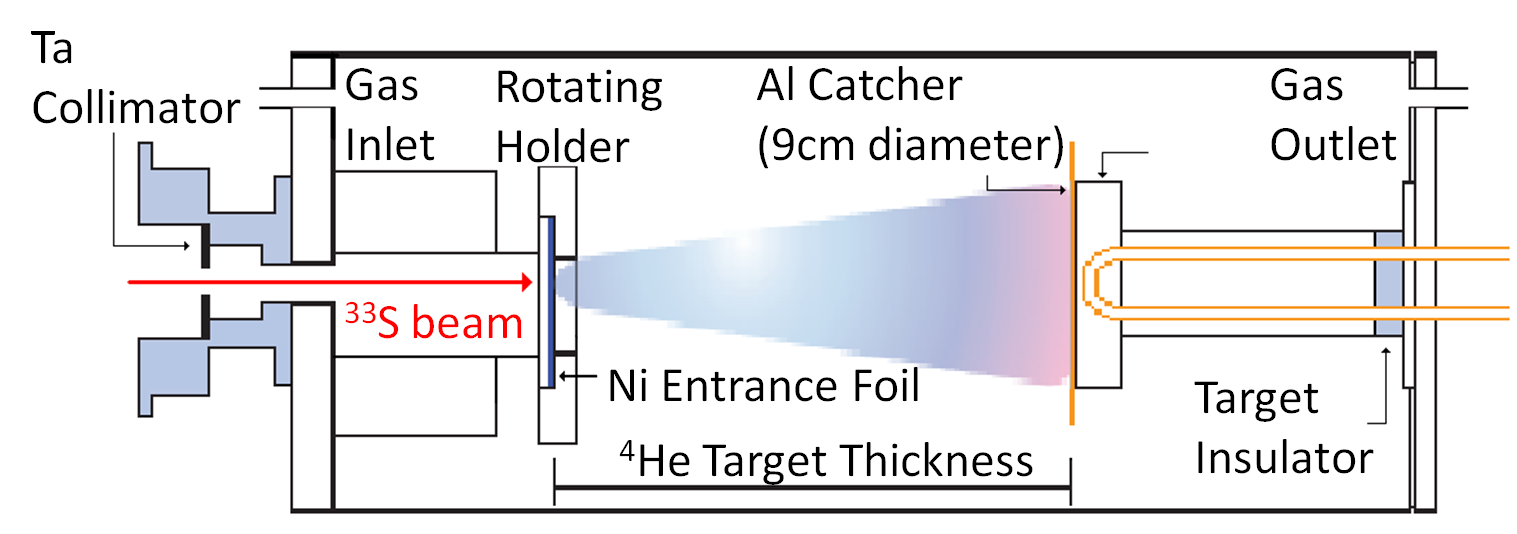}
\caption{\label{fig:GasCell}(color online) The gas cell used during activations. The insulator electrically isolated the gas cell from rest of beamline. A 9 cm diameter aluminum foil was used to catch the forward-recoiled \cl atoms. The target holder could be isolated from the rest of the chamber and used as a Faraday cup during beam tuning with the target insulator. The target insulator was then removed during activations and the entire gas cell was used as a Faraday cup.}
\end{figure}

The activations were performed at the  Nuclear Science Laboratory at the University of Notre Dame. A \iso{33}{S} beam was extracted from an FeS cathode, sent through the facility's 11 MV FN tandem accelerator, and focused on the gas cell. The chamber (Fig. \ref{fig:GasCell}) was electrically insulated from the beamline to allow total charge integration on target. A 2.5 $\mu$m thick Ni entrance foil was rotated during the activations to reduce beam-induced degradation of the foil. A target holder placed 24 cm behind the entrance foil acted as a beam stop. A 0.25 mm thick, 10$\times$10 cm$^2$ aluminum foil was attached to the front of the target holder during the activations to catch recoiled \cl atoms. The target holder was isolated from the rest of the gas cell during beam tuning with the target insulator and removed for the activations, which electrically connected the target holder to the rest of the chamber. The pressure in the gas cell during activations was 10 Torr, maintained by continuous \iso{4}{He} flow.

Before each sample activation, a beam current integration was performed with the target insulator installed and Ni entrance foil removed (see Fig. \ref{fig:GasCell}) and normalized to the current measured on a Faraday cup located $\sim$~0.5 m upstream, to monitor any changes in source output. A second beam current integration was performed with target insulator removed and the Ni entrance foil in place and again normalized to the current measured on the upstream Faraday cup. The agreement between the two measurements was between $\sim$1-6\% for all activation energies. This beam-tune dependent charge collection of the gas cell was included in the incident \3S ion flux uncertainty. An additional 2\% uncertainty was included in the number of incident \3S ions from the lab Faraday cups and gas cell readouts.

The activation times, average electrical beam current on target, and integrated number of incident \iso{33}{S} ions for each sample are summarized in Table~\ref{tbl:activation}. Sample S5 was previously activated and measured as a proof of principle sample~\cite{Bowers2012}. This sample's activation procedure was identical to that of the samples discussed in this paper. 

\begin{table}
\caption{\label{tbl:activation}Integrated beam current data for the \cl activations.}
\begin{ruledtabular}
\begin{tabular}{rrrr}
Sample			&	t (hr)	\footnotemark[1]	&	I$_{\text{avg}}$ (nA)\footnotemark[2]	&	N$_{33}$	($10^{14}$)\footnotemark[3]	\\
\hline
S1				&	43.16				&	104.4							&	145(12)				\\
S2				&	19.39				&	72.4								&	39.5(12)				\\
S3				&	5.97					&	68.5								&	13.1(5)				\\
S4				&	1.92					&	94.9								&	5.11(14)				\\
S5\footnotemark[4]	&	77.71				&	37.3								&	724(41)				\\
S6				&	2.98					&	12.4								&	0.83(5)				\\
\end{tabular}
\end{ruledtabular}	
\footnotetext[1]{Irradiation time.}
\footnotetext[2]{Average electrical beam current on target.}
\footnotetext[3]{Integrated number of incident \iso{33}{S} ions.}
\footnotetext[4]{Previously activated sample \cite{Bowers2012}.}
\end{table}	

To determine the energy range of the measured cross sections, the energy loss in the nickel entrance foil was experimentally determined, while the energy loss in the \iso{4}{He} target was calculated with the SRIM program \cite{SRIM}. To measure the energy loss in the nickel foil, the energy spectrum of the beam was measured with and without the nickel foil on a Si detector. The measurements were repeated for each activation energy.

\begin{figure}
\includegraphics[width = \columnwidth]{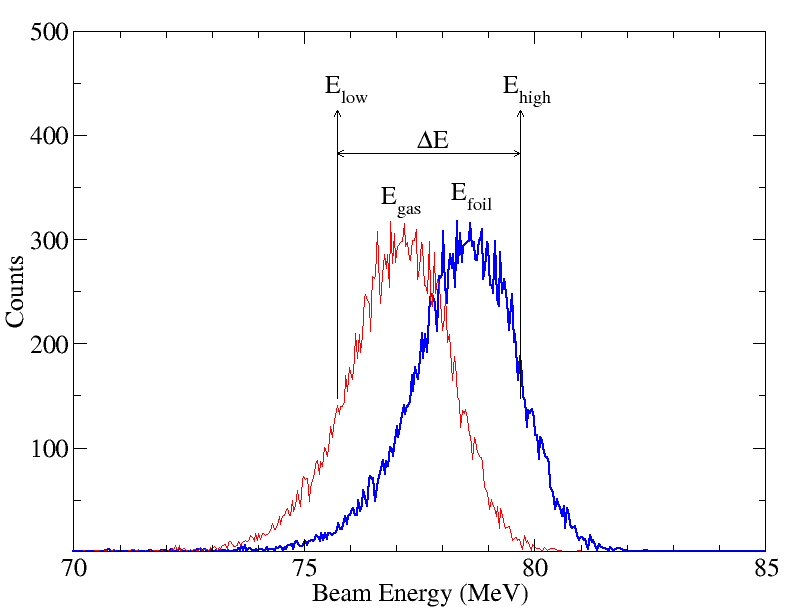}
\caption{\label{fig:Eloss} (color online) An example (sample S6) of the measured and calculated energy loss of \iso{33}{S} ions in the gas cell. E$_{\text{foil}}$ and  E$_{\text{gas}}$ are the centroids of the beam energy after the Ni entrance foil and \iso{4}{He} gas, respectively. The thick (blue) curve is the energy spectrum of the \iso{33}{S} ions after passing through the nickel foil. E$_{\text{high}}$ is the high energy end of the FWHM of the peak. The thin (red) curve is the bold curve shifted lower in energy by 1.44 MeV, the calculated energy loss in the \4He gas from SRIM, assuming constant stopping power over the energy spread. E$_{\text{low}}$ is the low energy end of the FWHM of the shifted peak. The derived cross sections are thus averaged over the energy range $\Delta$E.}
\end{figure}

The energy range for each activation was determined to be $E_{low}$ to $E_{high}$:

\begin{equation}
E_{high} = E_{foil} + \frac{FWHM}{2}
\label{eqn:Ehigh}
\end{equation}

\begin{equation}
E_{low} = E_{gas} - \frac{FWHM}{2},
\label{eqn:Elow}
\end{equation}
where $E_{foil}$ and $FWHM$ are the centroid and full width at half maximum, respectively, of the beam energy measured after the Ni entrance foil. $E_{gas}$ is the energy of the beam after the \iso{4}{He} gas target, evaluated by subtracting the energy loss in the gas calculated with SRIM from $E_{foil}$. The stopping power of the \iso{4}{He} gas was assumed to be constant through the gas cell because of the low pressure used and small beam currents on target ($<$ 100 nA, see Table~\ref{tbl:activation}). The cross sections are an average over the energy range $\Delta E = E_{high} - E_{low}$. There is a $\sim$20\% ($\sim$0.2 MeV) uncertainty of the calculated stopping power of heavy ions in gases using SRIM. However, the SRIM uncertainty is an order of magnitude lower than the measured energy spread of the beam, which dominates the energy range. An example energy loss measurement is shown in Fig. \ref{fig:Eloss}, with the results summarized in Table \ref{tbl:Eloss}.

\begin{table}
\caption{\label{tbl:Eloss}Results of the measured and calculated \iso{33}{S} ions' energy loss in the \iso{4}{He} gas cell. All energies are given in MeV.}
\begin{ruledtabular}
\begin{tabular}{rrrrrrrr}
Sample & E$_{\text{i}}$\footnotemark[1]		&	FWHM\footnotemark[2] 	&	E$_{\text{foil}}$\footnotemark[3] 	&	E$_{\text{gas}}$\footnotemark[4] 	&	E$_{\text{high}}$\footnotemark[5]	&	E$_{\text{low}}$\footnotemark[6]	&	$\Delta$E\footnotemark[7] \\
\hline
S1				&	56		&	2.4		&	26.2			&	24.3			&	27.4			&	23.1		&	4.3		\\		
S2				&	63		&	2.6		&	33.3			&	31.5			&	34.6			&	30.2		&	4.4		\\
S3				&	72		&	2.6		&	43.1			&	41.4			&	44.4			&	40.1		&	4.3		\\
S4				&	81		&	2.6		&	52.9			&	51.3			&	54.2			&	50.0		&	4.2		\\
S5\footnotemark[8]	&	90		&	2.7		&	62.8			&	62.0			&	64.2			&	60.7		&	3.5		\\
S6				&	104.5	&	2.8		&	78.4			&	77.0			&	79.8			&	75.6		&	4.2		\\
\end{tabular}
\end{ruledtabular}
\footnotetext[1]{The initial \iso{33}{S} beam energy before the gas cell entrance foil.}
\footnotetext[2]{The FWHM of the \iso{33}{S} beam after the Ni entrance foil.}
\footnotetext[3]{The mean energy of the \iso{33}{S} beam after the Ni entrance foil.}
\footnotetext[4]{The mean energy of the \iso{33}{S} beam after the \iso{4}{He} gas target calculated with SRIM.}
\footnotetext[5]{The high end of the activation energy range calculated with Eqn. \ref{eqn:Ehigh}.}
\footnotetext[6]{The low end of the activation energy range calculated with Eqn. \ref{eqn:Elow}.}
\footnotetext[7]{$\Delta$E = E$_{\text{high}}$ - E$_{\text{low}}$ = cross-section energy range.}
\footnotetext[8]{Previously activated sample \cite{Bowers2012}.}
\end{table}

\begin{figure}
\includegraphics[width = \columnwidth]{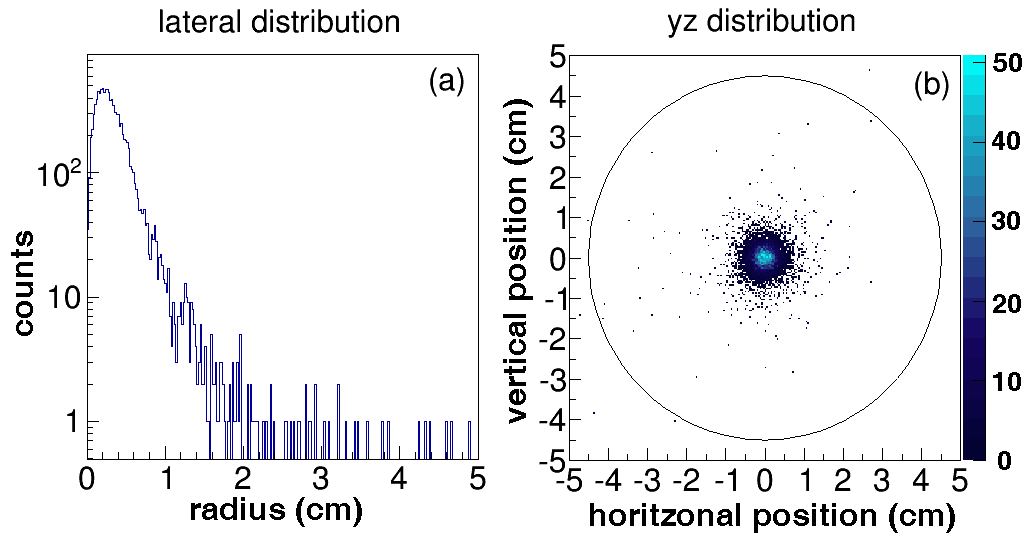}
\caption{\label{fig:lateral}(color online) TRIM simulation of $10^4$ \iso{33}{S} ions through the \iso{4}{He}-filled gas cell for sample S6. Figure (a) is a histogram of the lateral distribution of the implanted ions. The beam, before dispersion, is centered at a radius = 0. Figure (b) is a 2-D histogram of the particles implanted into the aluminum catcher foil. The circle is the 9 cm diameter opening of the Al catcher foil. }
\end{figure}

To ensure the \6Cl atoms were collected in the catcher foil, a TRIM simulation was used to determine the dispersion of the beam through the gas cell. The simulation tracked $10^4$ \iso{33}{S} ions passing through the Ni entrance foil, 10 Torr of \iso{4}{He} gas (5 Torr for sample S5), and embedded in the Al catcher foil. For all energies, greater than 99\% of ions were collected within the Al catcher foil's 9 cm diameter opening. An example of one of the simulations is shown in Fig. \ref{fig:lateral}.

\subsection{Chemical Processing}
\label{sec:chemistry}

The Al catcher foils were chemically processed at Purdue University's PRIME lab \cite{Vogt1994}. In addition to the 5 activated samples, 2 identical, but non-irradiated foils were processed as blanks for the AMS measurement.  The foils were cut into 8 pieces, put in separate containers, and mixed with stable chlorine carrier (1.101 mg/g chlorine concentration), where the precise Cl-carrier masses for each sample is given in Table \ref{tbl:chemistry}. The addition of the Cl carrier fixes the \ratiocl ratio in each sample since both the \cl and stable chlorine are recovered with the same efficiency. Twenty mL HNO$_3$ (trace metal grade, 70\% concentration) were added to each sample to dissolve the foils. Since aluminum oxidizes in nitric acid, which inhibits its dissolution, 10 mL of HF (40\% concentration) were added to prevent oxidation. Then, 45 mL of 18 M$\Omega$ DI H$_2$O were added to slow down the reaction. After one hour, an additional 20 mL of HNO$_3$ were added to the samples and left overnight to dissolve. Before decanting the solution in separate vials, 10 drops ($\sim$0.5 mL) of AgNO$_3$ were added to precipitate the Cl as AgCl and the aliquots were centrifuged. The excess solution was decanted, leaving behind the precipitated AgCl. The samples were finally baked for 2 days at 70\degree~C to remove any excess moisture. Since the AMS system at the University of Notre Dame can separate \cl from its stable isobar, \iso{36}{S}, there was no need to chemically reduce the sulfur in the samples (section \ref{sec:ams_measurement}). Sample S5 was chemically processed in a similar protocol to that described above \cite{Bowers2012}.

The number of chlorine carrier atoms added to each sample ($N_{\text{Cl}}$) was calculated by

\begin{equation}
N_{\text{Cl}} = \frac{m_{carrier} \times \left( 1.101 \text{ mg/g} \right) \times N_A}{M_{\text{Cl}} \times 1000 \text{ mg/g}},
\label{eqn:NCl}
\end{equation}
where $m_{carrier}$ (g) is the mass of the Cl carrier added to the sample, $N_A$ (=$6.02\times10^{23}$ atoms/mol) is Avagadro's number, and $M_{\text{Cl}}$ (=35.4527 g/mol) is the atomic weight of chlorine. The carrier mass was multiplied by 1.101 mg/g to arrive at the mass of chlorine added to each sample. The Cl carrier mass and number of atoms added to each sample are given in Table \ref{tbl:chemistry}. The uncertainty in carrier mass, chlorine concentration, and chlorine recovery is estimated at 1\%.

\begin{table}
\caption{\label{tbl:chemistry}The carrier mass and number of Cl atoms added to each sample.}
\begin{ruledtabular}
\begin{tabular}{rrr}
Sample							&	m$_{\text{carrier}}$ (g)	&	N$_{\text{Cl}}$ ($10^{20}$)		\\
\hline
S1								&	12.4643				&	2.33(2)						\\
S2								&	12.4309				&	2.32(2)						\\
S3								&	12.6888				&	2.37(2)						\\
S4								&	12.6008				&	2.36(2)						\\
S5\footnotemark[1]					& 	49.9923				&	9.35(9)						\\
S6								&	12.7232				&	2.38(2)						\\
Blank1							&	12.5357				&	2.34(2)						\\
Blank2							&	10.2885				&	1.92(2)						\\
\end{tabular}
\end{ruledtabular}
\footnotetext[1]{Previously activated sample \cite{Bowers2012}.}
\end{table}

\subsection{AMS measurement}
\label{sec:ams_measurement}

The \ratiocl ratio in the samples was measured with the AMS system at the University of Notre Dame \cite{Robertson2007,Robertson2008}. The system uses a converted Browne-Buechner spectrograph with a 1 m radius, single-dipole magnet. Ion position and energy loss are measured after the spectrograph with a parallel grid avalanche counter (PGAC) and ionization chamber (IC), respectively. The detector system is described in \cite{Robertson2008}.

\begin{figure}
\includegraphics[width = \columnwidth]{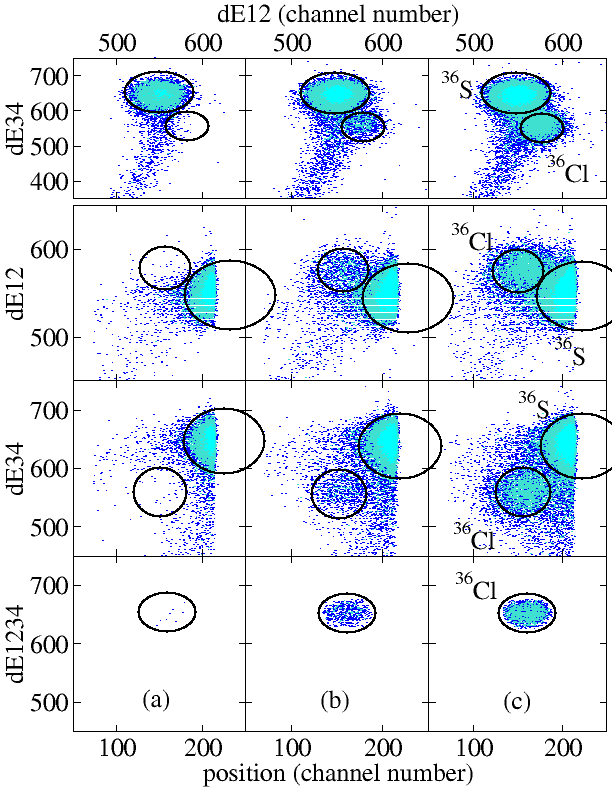}
\caption{\label{fig:spectraAll}(color online) Identification spectra of \cl.  The left column (a) is a 15 minute measurement of Blank1. The middle column (b) is a 15 minute measurement of activated sample S6. The right column (c) is a 10 minute measurement of the standard (\ratiocl = \sci{4.16}{11}). The top row shows the separation of the \cl and \iso{36}{S} groups in energy loss in the first two anodes (dE12) in the ionization chamber (IC), plotted versus energy loss the last two anodes (dE34). The second and third rows plot the position of the groups versus dE12 and dE34, respectively. The sharp cutoff of the \iso{36}{S} group is due to the beam shield, which blocked the majority of the \iso{36}{S} beam from entering the detector. The bottom row shows spectra of position versus total energy loss(dE1234)  in the IC. The spectra are gated on the \cl groups in the top three rows' spectra. The counts in the \cl group from the blank sample come from the tail and spread of the much more intense \iso{36}{S} beam.}
\end{figure}

The difficulty in measuring \cl arises from the need to separate it from its stable isobar, \iso{36}{S}. To separate \cl from \iso{36}{S}, we used the Gas- Filled Magnet (GFM) approach \cite{Paul1989}, since conventional electro-magnetic beamline elements are unable to separate the two isobars. In the GFM, the \iso{36}{Cl} and \iso{36}{S} ions separated into two different atomic number-dependent mean charge state groups and are bent in the GFM with different radii. The resulting peaks can then be distinguished in the position-sensitive PGAC. To achieve this separation the spectrograph was filled with 2.3 Torr of N$_2$ gas, which was isolated from the rest of the beamline with a 350 $\mu$g/cm$^2$ Mylar window. Count rates are kept low by physically blocking the \iso{36}{S} beam from the detector with a movable shield. Figure \ref{fig:spectraAll} shows spectra of the standard, blank and an activated sample, where \cl is separated from \6S in both position and energy. A more detailed discussion of the detector settings can be found in \cite{Bowers2012}. 

\begin{figure}
\includegraphics[width = \columnwidth]{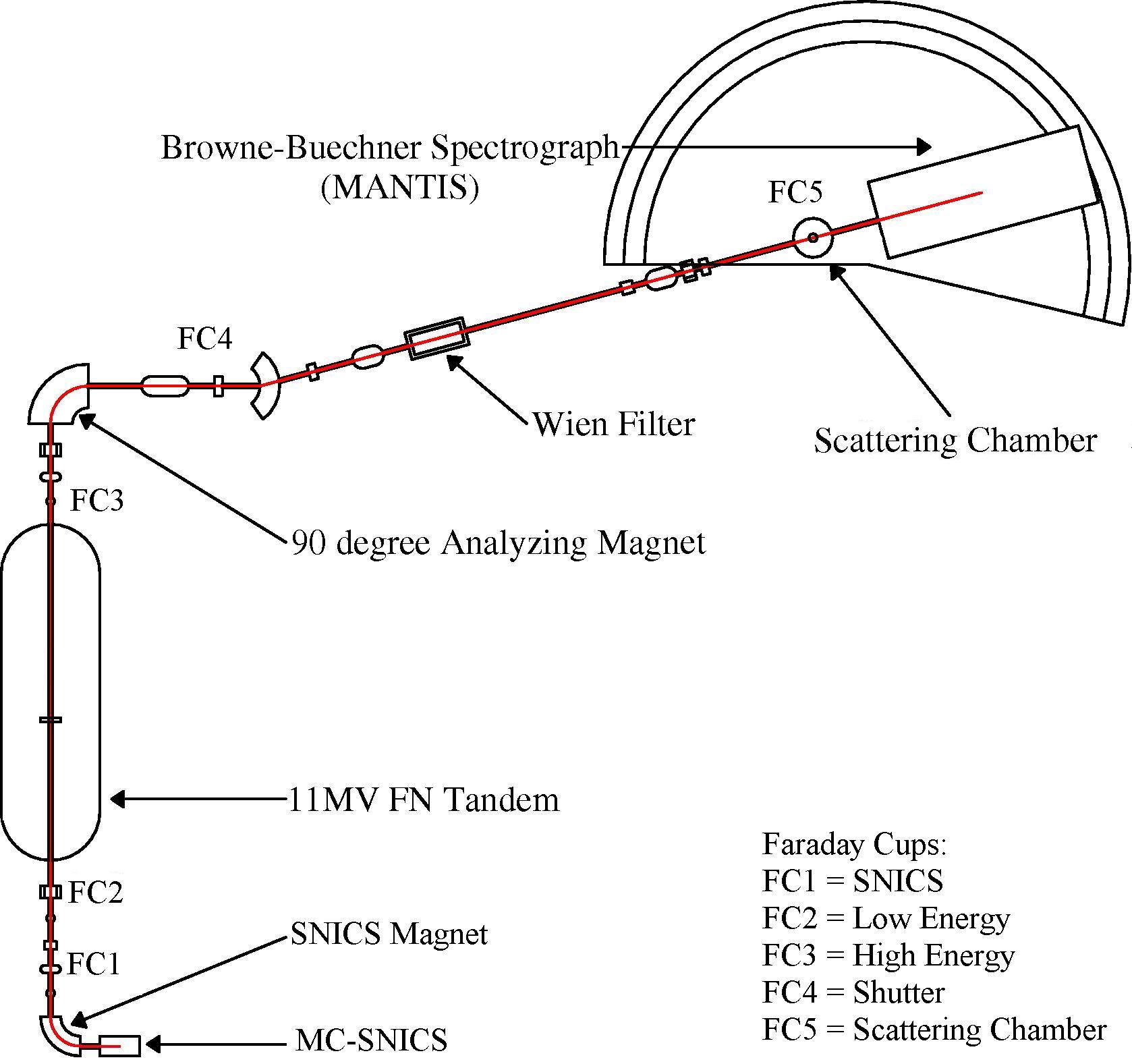}
\caption{\label{fig:beamline}Diagram of the AMS beamline.}
\end{figure}

\begin{figure}
\includegraphics[width = \columnwidth]{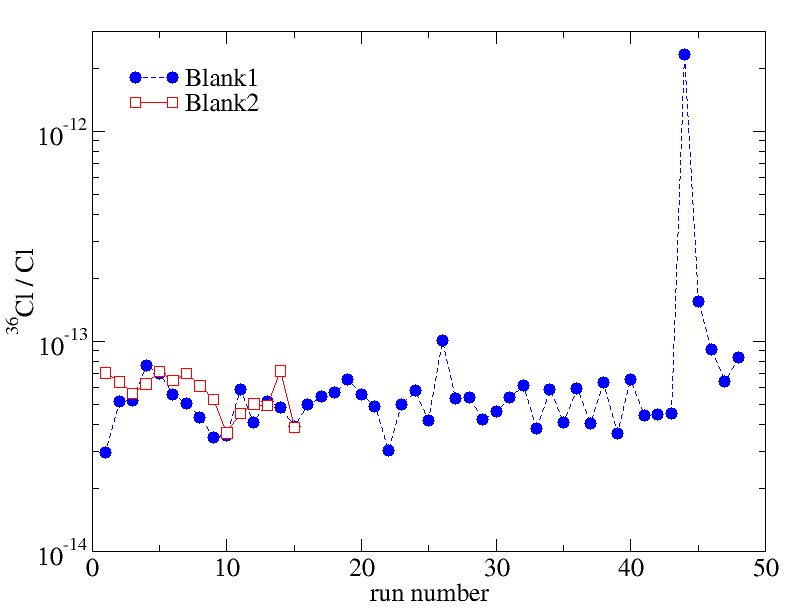}
\caption{\label{fig:blanks}(color online) Background measurements performed with Blank1 (circles) and Blank2 (squares). Yields of \cl for the activated samples were corrected for blank levels measured before each activated sample. Memory effects are seen in the spike at run 44, as it was the first blank measurement after the standard (\ratiocl = \sci{4.16}{11}) was measured. The \cl count rate decreased back to normal background levels after $\sim$1 hr.}
\end{figure}

The AMS measurement was performed with a Cl beam energy of 74.7 MeV and 8$^+$ charge state. The blank samples were measured multiple times to determine the background level and detection limit. The samples were then measured in multiple and independent measurements in order of increasing predicted \cl concentration to limit any potential source memory effects. The Blank1 sample was measured in between each measurement of activated samples. Blank2 verified the background levels determined with Blank1. The \iso{35}{Cl} beam current was recorded on Farady cup 1 (FC1) (See Fig. \ref{fig:beamline}), before and after each  \cl counting measurement, to normalize yields to source output. A summary of the blank measurements is shown in Fig. \ref{fig:blanks}. Individual measurement times varied from 10 to 60 minutes, depending on \cl concentration and source output. 

The transmission was measured with the \iso{35}{Cl} beam currents on Faraday cups FC1 to FC5 (Fig. \ref{fig:beamline}) between each sample measurement ($\epsilon_{35}$). This required reducing the beam output at the ion source ($<$ 1$\mu$A) before sending the beam through the accelerator. The transmission was also measured with \cl in the detector with a standard, \ratiocl = \sci{4.16}{11}, and accounted for beamline and gating losses, and detector efficiency ($\epsilon_{36}$). The \6Cl standard was obtained from Prime Lab and was originally prepared from an aliquot of a dilution series of NBS SRM 4422L~\cite{Vogt1994}. An uncertainty of 2\% is assigned to the standard from uncertainty in the original reference material activity and subsequent AMS measurements of the standard~\cite{Vogt1994}. The \ratiocl ratio of the activated samples was normalized to the standard. For each sample, the transmission measured with the standard ($\epsilon_{36}$) was scaled to the transmissions measured with the \5Cl beam between sample measurements ($\epsilon_{35}$) by 

\begin{equation}
\label{eqn:trans}
\epsilon_{36,sample} = \epsilon_{36,standard} \left( \frac{\epsilon_{35,sample}}{\epsilon_{35,standard}} \right).
\end{equation}
The statistical variation in the standard measurements to obtain the transmission was 4\%.

\begin{figure}
\includegraphics[width = \columnwidth]{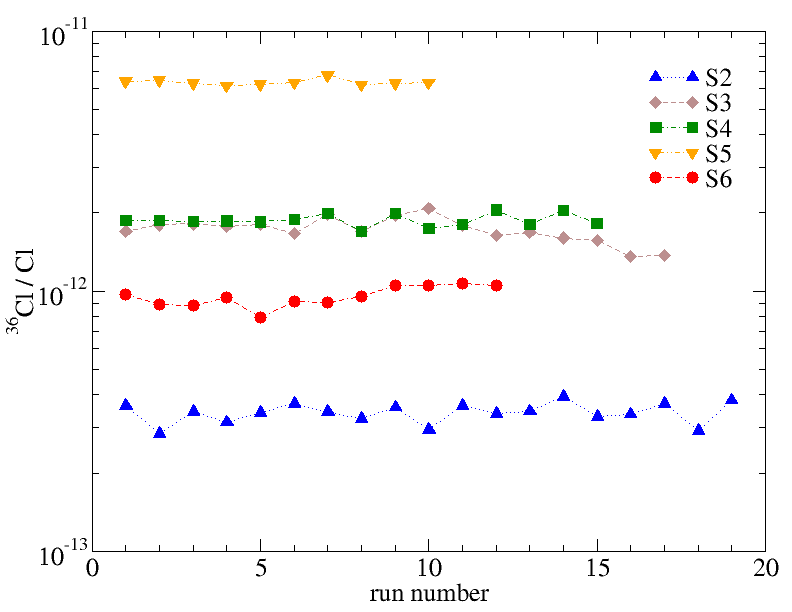}
\caption{\label{fig:AMS}(color online) A summary of the activated sample measurements: S2 (upright triangles), S3 (diamonds), S4 (squares), S5 (upside down triangles), and S6 (circles). Sample S1 showed no excess \cl above the blank level and is not shown.}
\end{figure}

The \ratiocl measurements are summarized in Fig. \ref{fig:AMS} and Table \ref{tbl:results}. The \ratiocl values are the unweighted mean of the different measurements of each sample. The uncertainty is given as one standard deviation of the mean. The previously activated and measured sample, S5, was remeasured to obtain a more accurate result and used as a test for the reproducibility of the AMS measurement. The remeasured value of S5, \ratiocl = $(6.4 \pm 0.3) \times 10^{-12}$, is in excellent agreement with the previously measured value of $(6.2 \pm 1.1) \times 10^{-12}$ \cite{Bowers2012}. Sample S1 showed no excess of \cl above the blank level, so its result is quoted as an upper limit.

\begin{table}
\caption{\label{tbl:results}Results of the experimentally determined average cross sections.}
\begin{ruledtabular}
\begin{tabular}{lccrr}
Sample & E$_{\text{low}}$ - E$_{\text{high}}$\footnotemark[1]		& \ratiocl 	& N$_{^{36}\text{Cl}}$ 			&	$\left< \sigma \right>$   \\
	   &		(MeV/A)									& 		& ($10^8$)					& (mb) \\
\hline
S1	&	0.70 - 0.83								& 	$5 	  \times 10^{-14}$\footnotemark[2] 	& 0.12\footnotemark[2]  	& 	0.1\footnotemark[2]  \\
S2	&	0.92 - 1.05								&	$3.4(3) \times 10^{-13}$  			& 0.79(8) 						& 	2.4(3) \\
S3	&	1.22 - 1.35 								& 	$1.7(2) \times 10^{-12}$ 				& 4.0(5) 						& 	37(5) \\
S4	&	1.51 - 1.64 								& 	$1.9(1) \times 10^{-12}$ 				& 4.5(3) 						& 	105(8) \\
S5	&	1.84 - 1.95 								& 	$6.4(3) \times 10^{-12}$ 				& 59.8(32) 					& 	199(16) \\
S6	&	2.29 - 2.42								& 	$9.6(9) \times 10^{-13}$ 				& 2.3(2) 						& 	330(40) \\
\end{tabular}
\end{ruledtabular}
\footnotetext[1]{Energy is converted from MeV to MeV/A. Values from table \ref{tbl:Eloss} were divided by 33, the atomic number of \iso{33}{S}.}
\footnotetext[2]{Upper limit.}
\end{table}

\section{Results}
\label{sec:results}

The cross section was determined by 

\begin{equation}
\left< \sigma \right> = \frac{N_{^{36}\text{Cl}}}{N_{33} \times N_T},
\label{eqn:sigma}
\end{equation}
where $N_{33}$ is the total number of incoming \iso{33}{S} ions during the activation (Table \ref{tbl:activation}). The number of \cl atoms in the sample ($N_{^{36}\text{Cl}}$) was found by multiplying N$_{\text{Cl}} \times ^{36}\text{Cl}/\text{Cl}$, determined from the chemical processing and AMS measurement, respectively. The area density of the \iso{4}{He} target atoms ($N_T$) is given by

\begin{equation}
N_T = \rho_{atm}\left(\frac{P}{P_{atm}}\right)\left(\frac{N_A}{M_{\text{He}}}\right) d,
\label{eqn:N}
\end{equation}
where $N_T$ is given in units of target nuclei/cm$^2$, $\rho_{atm}$ (=0.1664~g/cm$^3$) is the density of \iso{4}He at atmosphere, $P$ and $P_{atm}$ (Torr) are the pressure in the gas cell and atmospheric pressure, respectively. $M_{\text{He}}$ is the atomic weight of helium (=4.0026~g/mol) and $d$ (=24~cm) is the length of the gas cell from the Ni entrance foil to Al catcher foil. The experimentally determined cross sections are given in Table \ref{tbl:results} along with their associated energy ranges (now expressed in MeV/A). 

A summary of the uncertainties in the measurement is given in Table \ref{tbl:uncert}. 

\begin{table}
\caption{\label{tbl:uncert}Sources of uncertainty.}
\begin{ruledtabular}
\begin{tabular}{lrr}
								&	Statistical		&	Systematical			\\
\hline
Incident \3S ions (N$_{33}$)			&				&	1-6\% + 2\%			\\
Stable Cl carrier atoms (N$_{\text{Cl}}$)	&				&	1\%					\\
\4He target density					&				&	2.1\%				\\
AMS measurement					&				&						\\
Standard							&	4\%			&	2\%					\\
\6Cl/Cl							&	3-11\%		&	6\%\footnotemark[1]		\\
\end{tabular}
\end{ruledtabular}
\footnotetext[1]{From transmission and normalization to standard.}
\end{table}

\section{Discussion}
\label{sec:discussion}

\subsection{Comparison with Theoretical Predictions}
\label{sec:comparison}

\begin{figure}
\includegraphics[width = \columnwidth]{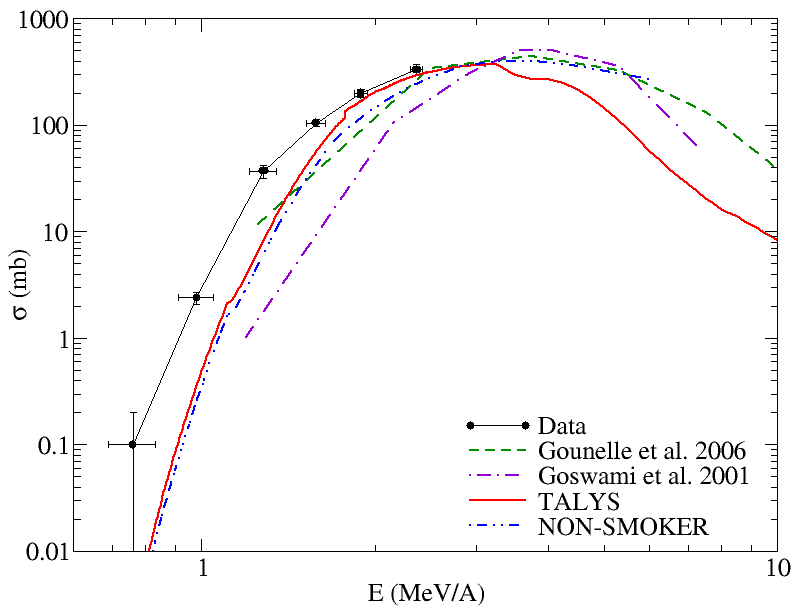}
\caption{\label{fig:plot}(color online) Comparison of the measured cross section and theoretical cross sections used in the ESS irradiation models from \cite{Gounelle2006} and \cite{Goswami2001}, and calculated with NON-SMOKER \cite{Rauscher2000,Rauscher2001} and TALYS \cite{TALYS,Koning2003}.}
\end{figure}

In order to evaluate the effect of the experimentally measured cross section on calculated \cl production in the Early Solar System, the experimental cross sections were compared to theoretical predictions, including those used in two ESS irradiation models \cite{Goswami2001,Gounelle2006}.  In addition, the cross sections were also compared to the statistical model codes TALYS (using the default parameters, see below) \cite{TALYS,Koning2003} and NON-SMOKER \cite{Rauscher2000,Rauscher2001}, that calculate cross sections with the Hauser-Feshbach model. The comparison of the experimental data to the theoretical cross sections are shown in Fig. \ref{fig:plot}. 

Over the measured energy range all of the theoretical predictions are lower than the data. The measured cross section is over an order of magnitude higher than that used in Goswami et al. 2001 at low energy ($<$2 MeV/A). The discrepancy improves with increasing energy yet fails to be resolved. The Gounelle et al. 2006 cross sections were also under predicted. However, the deviation between experiment and theory is resolved with the highest energy data point (sample S6). This would mean that these irradiation models under-calculated the \6Cl production via the \3S($\alpha$,p) reaction. Section \ref{sec:astro} discusses the potential effects of the experimental cross sections on \6Cl production.

Though differences between the results of various codes and the data tends to diminish with increasing energy, TALYS and NON-SMOKER predictions give, in general, a better description of the experimental excitation curve than the cross sections used in the irradiation models. Below $\approx$ 2.3 MeV/A there is a good agreement between TALYS and NON-SMOKER calculations, with differences not exceeding a factor of 1.25. At higher energies, where no experimental data exists, TALYS cross section predictions drop off more rapidly than other model calculations due to the inclusion of additional reaction channels.

The NON-SMOKER calculations were obtained using level densities given by the constant temperature plus back-shifted Fermi gas (CT+BSFG) model~\cite{Gilbert1965}, where the back shift and level density parameters are from~\cite{Rauscher1997}. In terms of the optical model potential (OMP), the NON-SMOKER calculations have been performed using the semi-microscopic neutron and proton OMP from \cite{jlm1977,Lejeune1980} (JLM), with an $\alpha$-OMP given by~\cite{MS1966} (MS). TALYS uses the CT+FG model as the default prescription for the level densities. To test the sensitivity of the TALYS results to the choice of level density model, identical calculations were performed using the CT+BSFG and BSFG models. Over the energy range covered by the experimental data, agreement between the two level density model calculations was found to be within 5\%. Similarly, identical calculations were also performed using two different optical model potentials (OMP). By default TALYS uses a phenomenological OMP, based on smooth energy-dependent forms for the potential depths, where widths and diffusenesses are from global averages \cite{Koning2003}. Results given by this model were compared to those obtained using the JLM model. The JLM results were found to be enhanced by at most 20\%. Cross section calculations were also performed using the MS $\alpha$-OMP. For energies below 1.6~MeV/A, the MS cross section predictions were reduced by approximately 20\%. Above this incident energy, differences between the MS and default OMP calculations become negligible.

In summary, it was found that over the measured energy range TALYS systematically under-predicts the experimental data, a finding that is not sensitive to either the OMP, $\alpha$-OMP, or level density model choice.

\subsection{\cl Production in the Early Solar System}
\label{sec:astro}

While previous studies of \cl in the early Solar System sought to reproduce the initial (\6Cl/\5Cl)$_0$ ratio inferred from meteorite measurements~\cite{Goswami2001,Leya2003,Gounelle2006}, here we examine the effects of using the measured \3S($\alpha$,p) cross section compared to using the theoretically predicted values, as well as investigate how the astrophysical environment parameters affect which reactions are most important to \6Cl production. The study was performed by adapting the irradiation model developed by Gounelle et al. 2001~\cite{Gounelle2001} and subsequently used by Gounelle et al. 2006~\cite{Gounelle2006}. To ensure the calculations were consistent with the previous studies the same parameters were used (see case 2d from~\cite{Gounelle2001}). This was the adopted case used by Gounelle et al. 2006 \cite{Gounelle2006}.

When considering the possible radiation emitted from the protoSun, the particles' energy spectrum and abundances must be established. The proton number flux was represented by a power-law distribution $\propto E^{-p}$, where $E$ is the proton energy in MeV/u and $p$ varies between 2.7 and 5. The \iso{4}{He}/\iso{1}{H} and \iso{3}{He}/\iso{1}{H} ratios scaled the proton number flux to give \iso{4}{He} and \iso{3}{He} fluxes. The Solar energetic particles (SEPs) originate from either impulsive (IMP) or gradual (GRD) events. Impulsive events are characterized by a sharper energy spectrum (larger $p$) and the presence of a \iso{3}{He} flux. Gradual events have a shallower energy spectrum (smaller $p$) and lack \iso{3}{He}. Three spectral parameter events (2 impulsive and 1 gradual) were taken from Gounelle et al. 2006~\cite{Gounelle2006} and one impulsive event from Leya et al. 2003~\cite{Leya2003}. The four event settings used in this study are summarized in Table \ref{tbl:event_params}.

\begin{table}
\caption{\label{tbl:event_params}Event parameters used in the irradiation calculations.}
\begin{ruledtabular}
\begin{tabular}{lrrr}
Event						&	p 	&	\iso{4}{He}/\iso{1}{H}		&	\iso{3}{He}/\iso{1}{H}	\\
\hline
IMP4\footnotemark[1]			&	4	&	0.1					&	0.3					\\
IMP5\footnotemark[1]			&	5	&	0.1					&	0.3					\\
IMPL\footnotemark[2]			&	4	&	0.05					&	0.05					\\
GRD\footnotemark[1]				&	2.7	&	0.1					&	0					\\
\end{tabular}
\end{ruledtabular}
\footnotetext[1]{Gounelle et al. 2006~\cite{Gounelle2006}}
\footnotetext[2]{Leya et al. 2003~\cite{Leya2003}}
\end{table}

\begin{figure}
\includegraphics[width = \columnwidth]{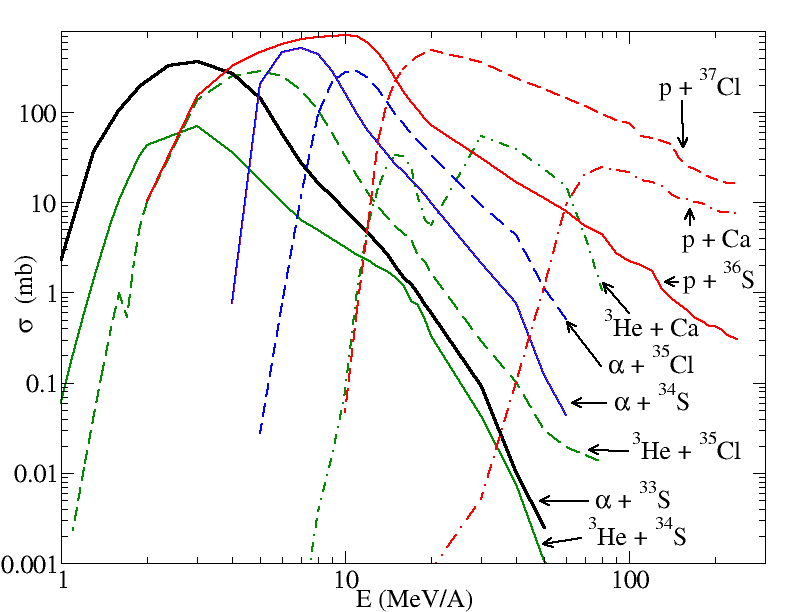}
\caption{\label{fig:xsPlot}(color online) The TALYS calculated cross sections for the \6Cl production reactions considered. The experimental data is included in the low energy sections of the \3S($\alpha$,p)\6Cl reaction.}
\end{figure}

The reaction cross sections on S, Cl, and Ca targets were calculated with TALYS with the default parameters for the code and are shown in Fig.~\ref{fig:xsPlot}. The experimental data ($<$2.4 MeV/A) is combined with the TALYS calculations for the \3S($\alpha$,p) reaction.

\begin{figure}
\includegraphics[width = \columnwidth]{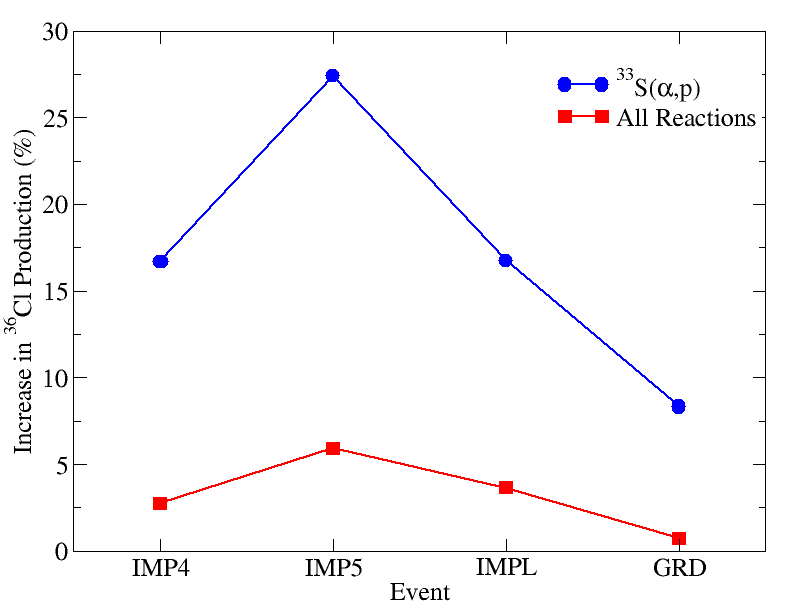}
\caption{\label{fig:data_effects}(color online) The increase in production of \cl via the \3S($\alpha$,p)\6Cl reaction as well as total production including all reactions using the experimental cross sections.}
\end{figure}

The calculations were performed with the TALYS cross sections and with the experimental data included with the TALYS predictions for all four events. The increase in production of \cl from the \3S($\alpha$,p) reaction along with the total increase in production, including all reaction channels, is shown in Fig.~\ref{fig:data_effects}. The effects are most dramatic for the IMP5 event where the particle flux is highest where the cross section has been measured. However, the increase in total production is $<7\%$ for all events. Although the measured cross sections are larger than theoretically predicted by as much as a factor of 3 for some energies, the overall effect on \cl production is minimal. As a check, the same calculations were performed with targets of a single chondritic elemental abundance. The results were consistent with the core-mantel composition since most of the particles are stopped in the mantle, where the volatile targets Cl and S are located. While the total effect of the measured cross sections on \cl are small, the results do show that the effect can vary depending on the event parameters. 

\begin{figure}
\includegraphics[width = \columnwidth]{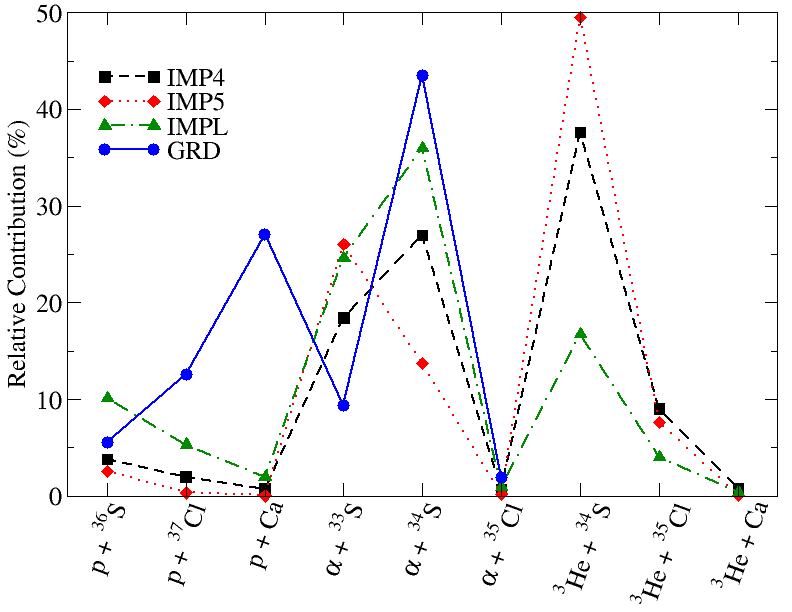}
\caption{\label{fig:relative}(color online) The relative contributions to the production of \cl of individual reaction channels for a particular event.}
\end{figure}

The effects of the various event parameters on the relative contributions of each individual reaction channel were tested, where the \cl produced via one reaction is divided by the total \cl produced for an event type. Fig.~\ref{fig:relative} shows the relative contributions of each reaction considered in the calculations. The most dominant channels for \cl production are via the $^{34}$S($^3$He,p), $^{34}$S($\alpha$,pn), and \3S($\alpha$,p) reactions. However the relative contributions of these reactions change substantially depending on the type of event considered. Reactions on $^{34}$S targets contribute more to \cl production than on \3S targets due to its larger isotopic abundance (4.21\% versus 0.75\%). In the absence of $^3$He and a shallower energy spectrum in the GRD event case the Ca(p,x) reactions start to contribute a substantial fraction of the \cl, which was not considered in~\cite{Gounelle2006}. In most cases, reactions on volatile targets like sulfur contribute the most to \cl production in the early Solar System.

\section{Conclusion}
\label{sec:conclusion}

We have presented the first experimental results for the \3S($\alpha$,p)\6Cl reaction. The cross section was measured at 6 energies between 0.70 - 2.42 MeV/A. The experimental results were shown to be lower than the theoretical predictions previously used in early Solar System irradiation models~\cite{Goswami2001,Gounelle2006}. The results were also compared to the TALYS and NON-SMOKER Hauser-Feshbach codes, where these calculations also under predict the experimental values. The OMP, $\alpha$-OMP, and LD model were varied using the TALYS code to try and resolve the discrepancy among the TALYS, NON-SMOKER, and the experimental data. However, it was found that the disagreements were not sensitive to those input models. Higher energy data would be useful to help resolve the discrepancies among the different theoretical models for this reaction.

The experimental cross sections increase the contribution from the \3S($\alpha$,p) reaction to \cl production but have a minimal effect on total \cl production. While the importance of the \3S($\alpha$,p) reaction was not as great as previously predicted, it was shown to be one of the dominant production reactions. The relative contributions of the important \6Cl-production reactions vary appreciably depending on the astrophysical event parameters. The results show the importance of reactions on volatile targets like sulfur, especially the $^{34}$S($^3$He,p), $^{34}$S($\alpha$,pn), and \3S($\alpha$,p) reactions. Currently the \3S($\alpha$,p) reaction is the only reaction of these experimentally measured. The TALYS predictions for the $^{34}$S($^3$He,p) and $^{34}$S($\alpha$,pn) reactions differ substantially from the cross sections calculated in~\cite{Gounelle2001}. Experimental investigation of these reactions would be important as the effects of these discrepancies are not trivial on \6Cl production. In a gradual event environment reactions on Ca contribute considerably to \cl production. The Ca(p,x) reactions have been experimentally measured~\cite{Schiekel1996,Sisterson1997,Imamura1997}.

\begin{acknowledgments}
The authors would like to thank Marc Caffee and Mike Bourgeois from PRIME Lab at Purdue University for providing the \6Cl standard material and for their guidance and help with the chemical preparation of the samples. The authors would also wish to express their gratitude to the staff of the Nuclear Science Lab at Notre Dame for their continued assistance with all technical aspects of this work. Lastly, the authors thank Michael Paul for his continued assistance with the development of the AMS system at Notre Dame. This work is supported by the National Science Foundation grant number NSF-PHY97-58100.
\end{acknowledgments}

\bibliography{bibliography}

\end{document}